# Probing optical and acoustic phonons in heated nano-Si/epoxy composites

Bayan Kurbanova[1,*], Vladimir Bessonov[1], Ivan Lysenko[2], Gauhar Mussabek[3], Ali Belarouci[4], Vladimir Lysenko[5], Yanwei Wang[6,7] and Zhandos Utegulov[1]

[1]Department of Physics, School of Sciences and Humanities, Nazarbayev University, Astana 010000, Kazakhstan
[2]Institute of High Technologies, Taras Shevchenko National University of Kyiv, 64/13 Volodymyrska St., Kyiv 01601, Ukraine
[3]Faculty of Physics and Technology, Al-Farabi Kazakh National University, Almaty 050040, Kazakhstan
[4]Lyon Institute of Nanotechnology, UMR 5270, INSA de Lyon, 69100 Villeurbanne, France
[5]Light Matter Institute, UMR-5306, Claude Bernard University of Lyon/CNRS, Université de Lyon, 69622 Villeurbanne Cedex, France
[6]Department of Chemical and Materials Engineering, School of Engineering and Digital Sciences, Nazarbayev University, Astana 010000, Kazakhstan
[7]Center for Energy and Advanced Materials Science, National Laboratory Astana, Astana 010000, Kazakhstan

*Corresponding author: bayan.kurbanova@nu.edu.kz

Understanding the thermal response of optical and acoustic phonons is crucial for designing functional polymer nanocomposites. We study silicon nanoparticle (Si NP)-epoxy composites (0.02–2 wt%) using combined Raman and Brillouin spectroscopy under *local* (laser-induced) and *global* (stage-controlled) heating. Raman spectra reveal THz longitudinal optical (LO) phonon softening and spectral broadening under local heating, indicating nanoscale hot-spots and interfacial scattering. Brillouin data track GHz longitudinal acoustic (LA) phonons, showing temperature- and concentration-dependent evolution of elasticity and damping. Contrasting heating methods unravels Si loading thresholds for isolated thermal absorbers, thermal percolation, acoustic attenuation and elastic homogenization. Local heating induces greater phonon softening and damping than global heating, with this disparity amplified at higher loadings by thermal gradients and interfacial dissipation. Global heating correlates with viscoelastic relaxation, showing intensified acoustic attenuation near the glass transition. Raman thermometry coupled with finite-element opto-thermal modeling allows evaluation of thermal conductivity of the composites characterized by increase from 0.09 to 0.46 W/(m·K) for 0.07 up to 2 wt% of Si NPs, respectively, outperforming SiC nanowires at 2 wt% [D. Shen et al, Sci. Rep. 7, 2606 (2017)] despite bulk conductivity of Si being more than 3 times smaller than that of SiC. However, effective heat conductivity of our nanocomposites remain far below bulk Si, confirming that interfacial thermal resistance, not filler conductivity, governs heat transport. Our results establish Si NP/epoxy as an *interface-engineered phonon-damping material with finely tuned thermal conductivity* and demonstrate multimodal spectroscopy's power to resolve thermo-phononic processes, with implications for vibration damping and thermal management materials.

Polymer nanocomposites have emerged as versatile functional materials for thermal management, sensing, flexible electronics, and structural applications due to their tunable thermal transport [1, 2] and mechanical properties [3 - 5] enabled by NP incorporation. These hybrid systems constitute an important class of multifunctional materials in which nanoscale fillers strongly influence heat transport, elasticity, and mechanical damping. Epoxy-based nanocomposites are particularly attractive for electronics packaging, thermal interface materials, structural adhesives, coatings, and vibration-damping components, where controlled heat dissipation and mechanical stability under thermal load are critical [6]. In such environments, NP loading, dispersion, and polymer–filler interfacial coupling govern both steady-state thermal conductivity and dynamic thermo-mechanical behavior under localized heat generation.

Thermosetting polymers are generally preferred over thermoplastics because of their higher stiffness and strength [7]. Among them, epoxy resins dominate owing to their excellent mechanical properties, strong adhesion to diverse reinforcements [8, 9], and broad industrial use in automotive, paint, and coating applications [10, 11]. However, epoxy exhibits limited impact strength and wear resistance, which are commonly improved through the incorporation of reinforcing fillers. Most epoxy nanocomposites rely on hard ceramic particles (e.g., $SiO_2$, SiC, $Al_2O_3$, WC) to enhance mechanical, tribological, and thermal performance [12]. In contrast, the thermo-phononic and viscoelastic behavior of Si NP-epoxy composites, particularly under localized versus uniform heating, remains largely unexplored.

Si NP-filled polymers are of particular interest because Si combines high intrinsic thermal conductivity, a well-defined phonon spectrum, and chemical compatibility with polymer matrices. Despite extensive efforts to enhance effective thermo-mechanical performance, a fundamental understanding of how optical and acoustic phonons respond to non-equilibrium heating is still lacking. Phonon-resolved studies that simultaneously probe THz optical and GHz acoustic dynamics under localized versus global heating conditions are rare, despite their relevance to realistic operating environments characterized by spatially heterogeneous heat generation.

Inelastic light scattering spectroscopies provide a powerful framework for addressing this gap by offering frequency-resolved access to lattice and mechanical dynamics that evolve with temperature(T) and microstructure. Raman scattering probes THz optical phonons, anharmonicity, interfacial strain, and local T rise, while Brillouin scattering accesses GHz acoustic phonons, yielding elastic moduli, sound velocities, and acoustic damping. Raman spectroscopy has been widely applied to crystalline Si and nanostructured systems [13 - 22], as well as to polymer nanocomposites [23]. Brillouin spectroscopy is a key tool for probing elastic and viscoelastic properties in polymers, nanocomposites, glasses, fluids, and biomaterials [3, 5, 24 - 40] and is increasingly recognized as complementary to Raman spectroscopy [41].

Previous studies often interpret Raman- or Brillouin- derived T responses as equivalent to equilibrium heating [42, 43], overlooking discrepancies between local and global thermo-phononic behavior in heterogeneous nanocomposites dominated by interfacial thermal resistance and viscoelastic losses.

In this work, we demonstrate for the first time a clear decoupling between local and global thermo-phononic responses in nano-Si/epoxy composites by combining Raman and Brillouin micro-spectroscopy. We identify distinct concentration-dependent thresholds governing isolated nanoheating, thermal percolation, acoustic damping, and elastic homogenization, and quantitatively link them to heat transport, phonon lifetimes, and viscoelastic relaxation.

A microwave plasma-assisted method decomposes a mixture of $SiF_4$ and $H_2$ to synthesize Si NPs with controlled size and high crystallinity. The reaction occurs in a ~600 cm³ quartz tube with a 2.45 GHz magnetron operating in single-pulse mode (1.6 kW, 5ms). Chamber pressure is held at 700 Torr, with an $H_2/SiF_4$ ratio of 6. The $SiF_4$ precursor also enables the formation of isotopically enriched Si NPs. The transmission electron microscopy (TEM) images and size distribution analysis of fabricated Si NPs are shown in FIG. 1(a). A schematic illustration of laser-induced (local) and stage-controlled (global) heating methods employed in our work is displayed in FIG. 1(b).

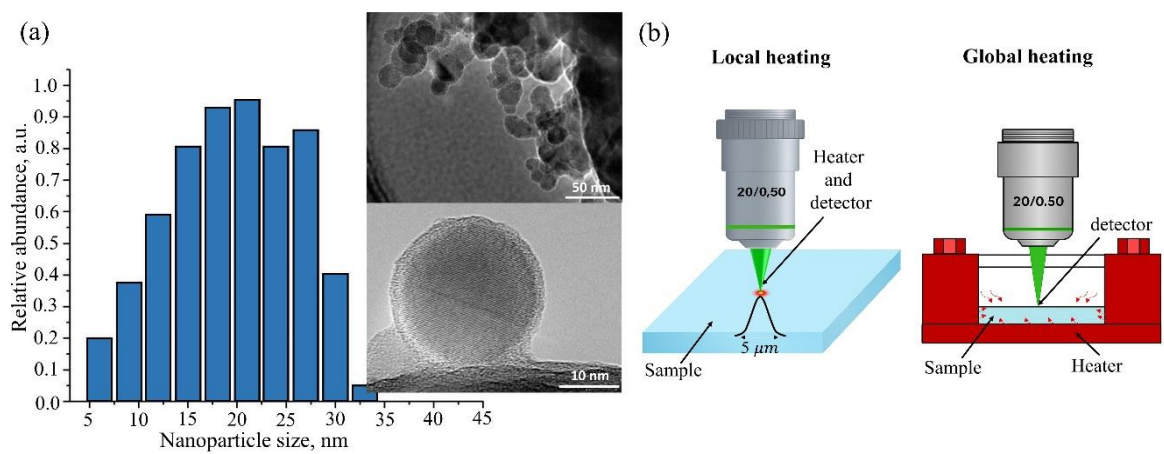

FIG.1. (a) Si NPs size distribution histogram, with low-magnification (upper inset) and high-magnification (lower inset) TEM images of the NPs, (b) Schematic illustration of local and global heating methods.

Optically transparent epoxy (Ultra Clear PRO, Epodex) serves as the matrix. The resin is mixed with a polyoxyalkyleneamine hardener in a 2:1 volumetric ratio, stirred for 15 minutes, and transferred to a 2.5 mL polyethylene container. A predetermined mass of Si NPs is placed in the container, followed by epoxy to reach 2 ml total volume. After 3 minutes of ultrasonic dispersion at 70% power using a probe sonicator, the mixture is placed on a rocking mixer for 48 hours to ensure uniform dispersion and curing. Samples are then aged for several days to complete internal structure formation.

Micro-Raman measurements with laser-induced local heating are performed using a continuous-wave, single longitudinal mode 532 nm diode-pumped solid-state laser (Laser Quantum Torus). Raman spectra are acquired using a Nanofinder 30 confocal Raman microscope coupled to a 35 cm spectrograph and an Andor iStar iCCD camera. A grating of 1800 grooves/mm provides a spectral resolution of ~1.4 cm$^{-1}$.

Photoinduced heat transport is simulated using finite-element modeling (FEM) implemented in Ansys Lumerical Heat. Steady-state Fourier heat conduction equations are solved using laser penetration depth values, with mesh sizes of 0.01–0.5 mm in the bulk and 0.01–0.1 μm near hot spots. Spatial T distributions are correlated with Raman thermometry measurements, with composite thermal conductivity as the sole fitting parameter. Evaluation of optical penetration depths is provided in the Supplementary Materials.

Micro-Brillouin measurements are conducted in a 180° backscattering geometry using a six-pass tandem Fabry–Perot interferometer (TFP-2, Table Stable Ltd.) with a free spectral range of 25 GHz and a 532 nm Verdi-G2 laser.

For Raman and Brillouin measurements under global heating the laser beam is focused to a ~ 5 µm spot using a 5× microscope objective, with incident powers ranging from 0.7 to 26 mW, adjusted according to Si loading. For global heating experiments, both Raman and Brillouin spectra are acquired using a TS1500 heating stage (Linkam). The positions, linewidths, and intensities of Raman and Brillouin spectral peaks are extracted via Lorentzian and Voigt fitting, respectively.

Studied Si NP-epoxy composites are getting progressively darker, as evidenced from the photographic images displayed in FIG. 2(a). Measured at ambient room-temperature Raman and Brillouin scattering spectra from LO Si-Si (520 $cm^{-1}$ = 15.6 THz) and LA (0.5 $cm^{-1}$ = 15 GHz) phonons of nano-Si/epoxy composites for different loadings of Si NPs at 5 mW and 9 mW incident laser powers are shown in FIG. 2(b) and FIG. 2(c), respectively. The Raman spectra clearly show how the chemical composition of the nanocomposite changes with the addition of Si. The characteristic (C-H, C-O-C and C=C) Raman peaks of the epoxy matrix progressively disappear with Si loading. The rapid dominance of the Si–Si Raman mode at low Si loadings arises from the exceptionally large Raman cross section of crystalline Si, resonance-enhanced scattering under visible excitation, and strong local-field and optical-weighting effects, causing Raman intensity to scale with optical polarizability rather than mass fraction.

At the same time, LO and LA phonon modes undergo a frequency downshift with the rise of NP loading [FIG. 2(b,c)] The addition of more Si into the matrix deteriorates the overall mechanical properties of the nanocomposite, attributed to the weak binding at the Si–epoxy interfaces. The detailed behavior of both phonon modes under Si loading at different laser-induced (local) and global (stage-controlled) temperatures is discussed with respect to FIG. 4 - 6 below.

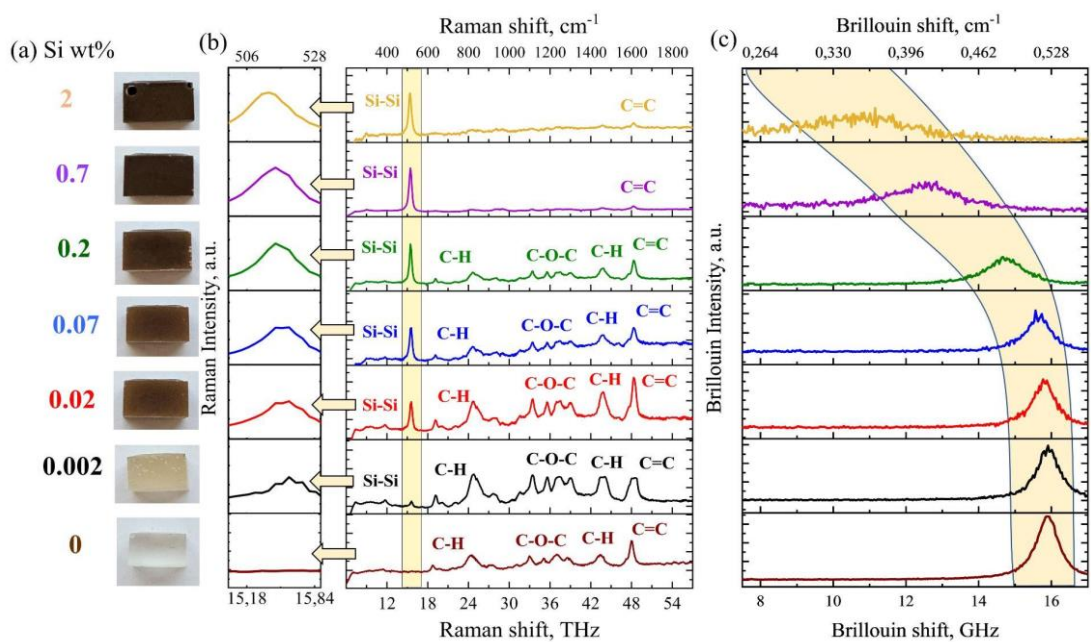

FIG.2. (a) Photographic images of Si NP–epoxy composites with increasing Si NP weight fraction (from bottom to top). Room-temperature (b) Raman and (c) Brillouin scattering spectra of Si NP–epoxy composites at different Si NP loadings, measured using incident laser powers of 5 mW (Raman) and 9 mW (Brillouin), respectively. The Si–Si longitudinal optical (LO) phonon (~15 THz) and longitudinal acoustic (LA) phonon (~15.9 GHz) modes are highlighted by shaded bands.

To determine heat conductive properties of the nanocomposites with Si loading, Raman thermometry coupled with opto-thermal FEM is employed. For this purpose local T under laser heating is determined by monitoring laser-induced Raman peak at 520 cm⁻¹ corresponding to the LO phonon mode of cubic Si NPs, as illustrated in FIG. 3(a,b). The Raman frequency shifts to lower frequencies and broadens as both local and global heating temperatures increase, while the C–H modes (Fig. 2b) attributed to the epoxy matrix exhibit no temperature-dependent Raman shift. The corresponding local T increase induced by the laser is estimated using the ratio of the integrated intensities of the Stokes and anti-Stokes Raman peaks [13]:

$$\frac{I_S}{I_{AS}} = \frac{(\alpha_I+\alpha_{AS})^3}{(\alpha_I+\alpha_S)} \frac{S(\omega_I,\omega_S)}{S(\omega_I,\omega_{AS})} e^{\frac{\hbar\omega_0}{k_B T}} \quad [1]$$

where, $I_S$ and $I_{AS}$ are the intensities of the Stokes and anti-Stokes bands at the same incident laser power. The ratio is proportional to the Boltzmann factor $e^{\frac{\hbar\omega}{k_B T}}$, $\omega_0$ - optical phonon frequency; $\omega_S, \omega_{AS}$ and $\omega_I$ are the Stokes, anti-Stokes and incident light photon frequencies, respectively; $\alpha_S$, $\alpha_{AS}$ and $\alpha_I$ are the optical absorption constants; $S(\omega_I,\omega_S)$ and $S(\omega_I,\omega_{AS})$ are the Raman cross-sections of inelastically scattered photons at Stokes and anti-Stokes sides, respectively.

Si loading-dependent photo-induced local temperatures in our nanocomposites are measured by Raman thermometry coupled with FEM modeling (FIG. 3a) and plotted as radial T distributions (FIG. 3(c-f)). Thermal conductivity is increased from 0.09±0.03 W/(m·K) to 0.46±0.07 W/(m·K) as Si NP loading rose from 0.07 to 2 wt% (FIG. 3, Table S1). This represents a ~3.4× enhancement, yet the nanocomposite's conductivity remains ~ 600× lower than bulk Si (145 W/(m·K)) and ~2.3× higher than neat epoxy (0.2 W/(m·K)) [2]. Heat transport is dominated by interfacial thermal resistance at the Si NP/epoxy interfaces, not by intrinsic Si conductivity, which is expected for ~20 nm NPs having a high surface-to-volume ratio. Phonon scattering at crystalline–amorphous interfaces is dominant [43].

With only 2 wt% filler, our enhancement matches the 0.449 W/m·K (106% increase) previously reported for epoxy/SiC nanowires (NWs) at 3.0 wt% and outperforms epoxy/SiC microparticles (MPs) at the same concentration [1]. This performance is particularly noteworthy given that the intrinsic thermal conductivity of bulk SiC exceeds that of bulk Si by more than a factor of three. Si NPs can act as highly effective thermal modifiers at low concentrations when interfacial effects are characterized. Further enhancement will require interface engineering, such as surface functionalization, to reduce Kapitza resistance [44] and better utilize Si's intrinsic conductivity.

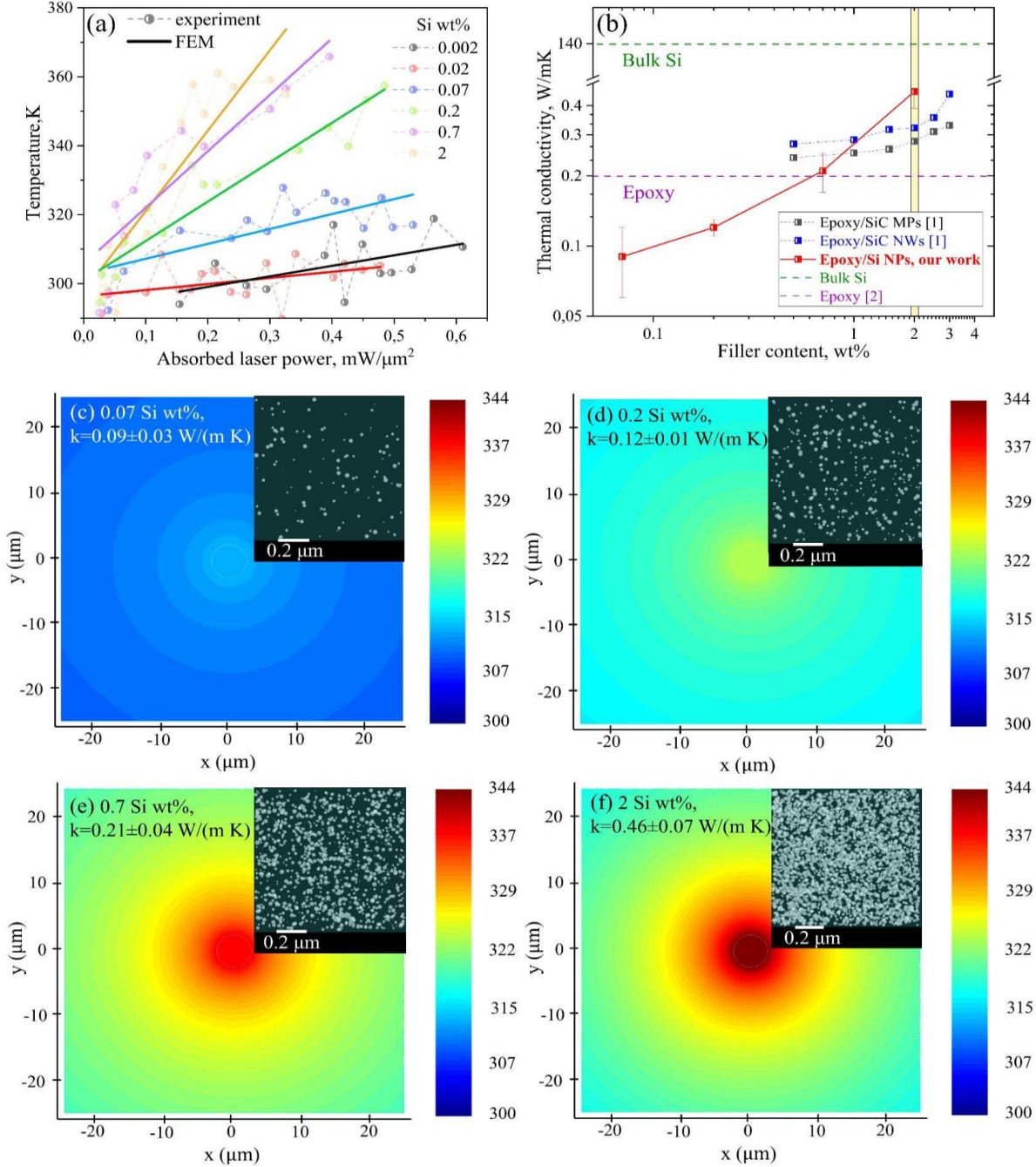

FIG. 3. (a) Local temperatures in np-Si/epoxy samples with various concentrations measured by Raman thermometry. (b) Thermal conductivities of epoxy-based nanocomposites versus filler content. FEM laser-induced radial T distributions and the corresponding thermal conductivity values of np-Si/epoxy samples with (c) 0.07, (d) 0.2, (e) 0.7 and (f) 2 wt%. The insets display randomly distributed NPs for each concentration.

Figure 4 summarizes LO and LA phonon evolution under laser-induced *local* heating versus Si NP loading. The Raman LO mode frequency $\omega_R$ downshifts with Si loading, especially at higher local temperatures [Fig. 4(a)]. The strongest changes occur between 0.05 and 0.2 wt%, saturating thereafter. This indicates tensile strain around NPs, enhanced phonon-phonon scattering, and epoxy network softening, showing Si NPs act as *thermo-elastic perturbation centers, not rigid reinforcements*. Concurrently, the Raman linewidth $\delta\omega_R$ increases monotonically with Si loading, particularly above ~0.1 wt% [Fig. 4(b)]. Since $\delta\omega_R \propto$ 1/phonon lifetime, this broadening reflects reduced optical phonon lifetimes from *enhanced anharmonic decay and interface scattering*. In contrast, the Brillouin LA mode frequency $\omega_B$ decreases sharply beyond ~ 0.05–0.1 wt%, showing strong T sensitivity exceeding that of the Raman shift [Fig. 4(c)]. This indicates a pronounced *reduction in effective elastic stiffness dominated by interfacial softening* [5]. The Brillouin linewidth $\delta\omega_B$ increases dramatically, up to ~5× with a sharp onset above ~0.1 wt%, especially at higher temperatures [Fig. 4(d)]. As $\delta\omega_B$ reflects acoustic damping, this broadening evidences strong interfacial friction and conversion of acoustic energy into heat. Overall, these nanocomposites exhibit *highly efficient phonon damping*.

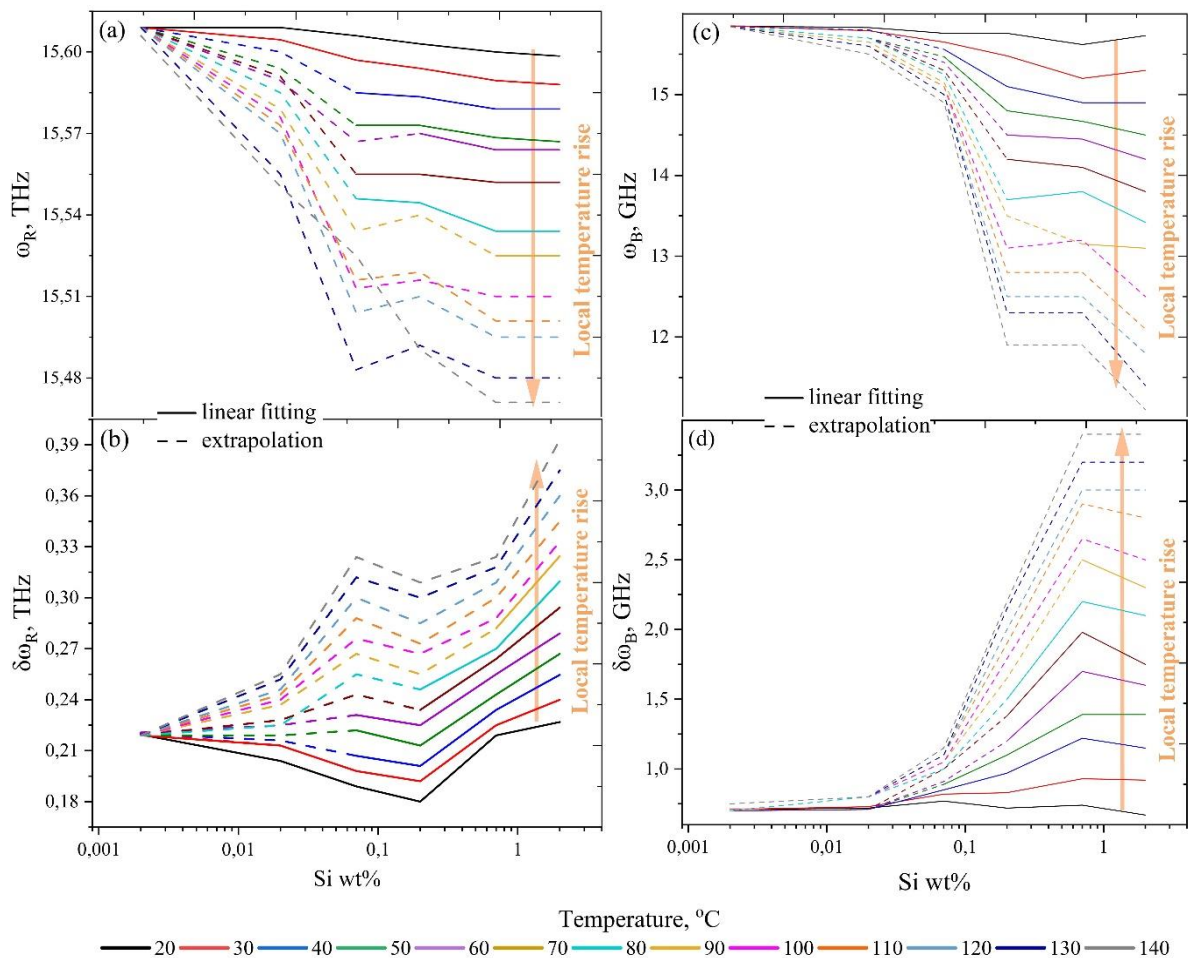

FIG. 4. Spectral shifts and linewidths of Raman peaks (a, b) and Brillouin peaks (c, d) versus different Si NP loading (wt%) at various *local* heating temperatures. Solid lines represent the linear fitting of measured results, while dashed lines show extrapolation of linear fitted data of Fig. S1 of the Supplementary Materials.

However, during *global* heating, a mixed behavior in the Si loading-dependent Raman and Brillouin frequency mode positions and linewidths is observed, as seen in Fig. 5.

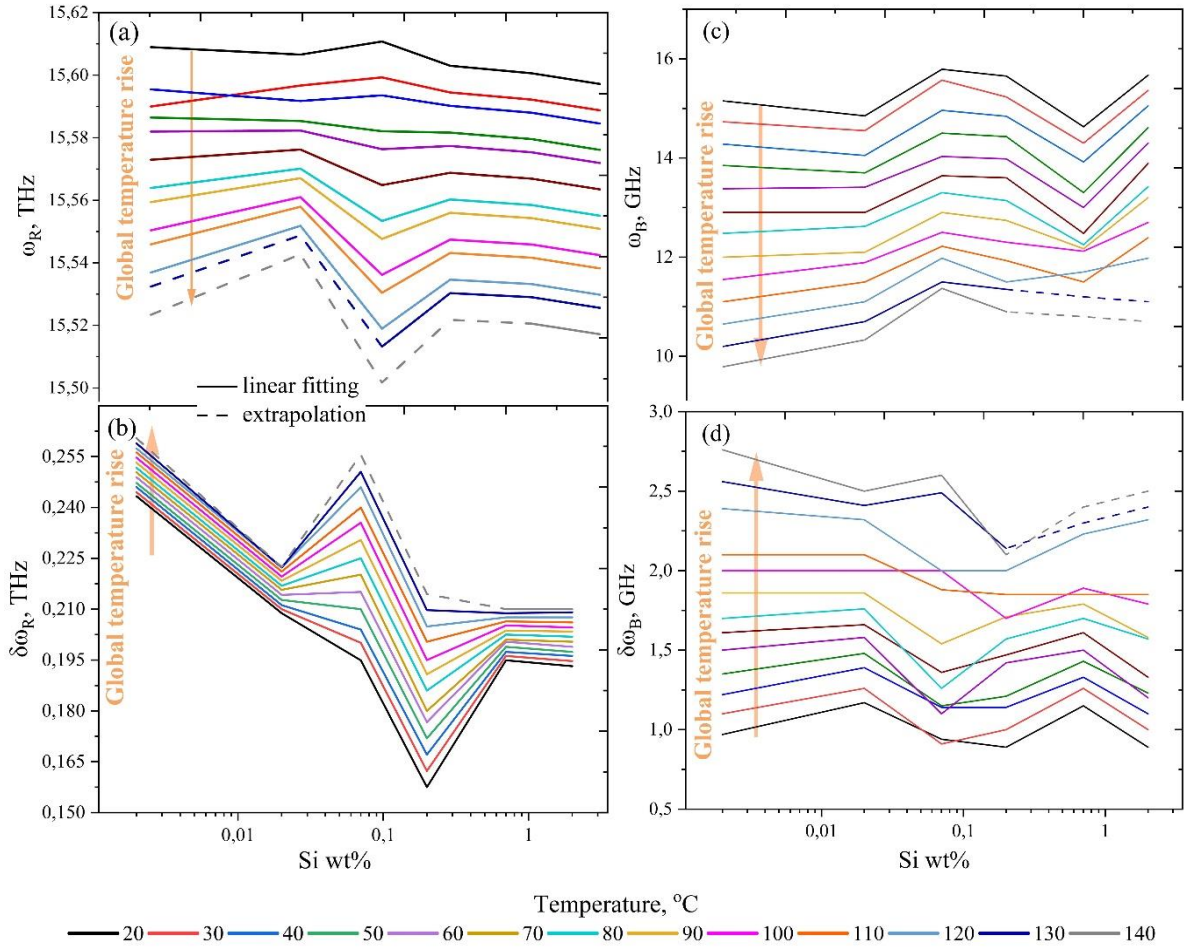

FIG. 5. Spectral shifts and linewidths of Raman peaks (a, b) and Brillouin peaks (c, d) versus different Si NP loading (wt%) at various *global* heating temperatures. Solid lines represent the linear fitting of measured results, while dashed lines show extrapolation of linear fitted data of Fig. S2 of the Supplementary Materials.

To distinguish between the effects of local and global heating, Fig. 6 presents the differences between local- and global heating-induced changes in Raman and Brillouin frequencies and linewidths as a function of Si loading at various global temperatures. Raman spectroscopy probes THz optical phonons and is highly sensitive to local T rises near optically absorbing Si NPs, whereas Brillouin spectroscopy probes GHz acoustic phonons, averaging over a larger volume and reflecting global elastic and viscoelastic responses. Plotting local–global differences isolates nonuniform heating, thermal transport, elastic homogenization, and damping mechanisms as Si loading increases. Increasing Si wt% enhances optical absorption, heat diffusion, elastic stiffness, and interfacial phonon scattering, giving rise to four distinct thresholds.

Figure 6(a) shows that local heating induces strong Raman linewidth broadening due to elevated local temperatures, enhanced anharmonic decay, and interfacial strain fluctuations, while global heating yields smaller linewidths owing to uniform T distribution. At very low Si loading, Si NPs act as isolated nanoheaters. At ~0.003 wt%, an isolated-absorber threshold is reached, marking the onset of absorption-dominated phonon damping before thermal interaction between NPs occurs. This localized hot-spot formation enhances phonon–phonon anharmonic decay and interface scattering relative to global heating [45].

Figure 6(b) shows that overlapping hot spots initially maximize Raman mode softening, yielding positive $\Delta_{R\ shift}$ because local heating generates higher nanoscale temperatures than global heating. As Si loading increases, heat diffusion pathways overlap and thermal percolation occurs, suppressing local T spikes. $\Delta R_{shift}$ crosses zero and becomes negative at ~0.07 wt%, defining the thermal percolation threshold where heat spreads efficiently through a percolated Si network. This behavior mirrors percolation transitions reported in epoxy composites filled with graphene and BN [46,47].

At low Si loading, local heating induces stronger viscoelastic losses than global heating, resulting in negative $\Delta_{B\ FWHM}$. With increasing Si content, polymer mobility is suppressed and damping becomes dominated by NP interfaces rather than T. $\Delta_{B\ FWHM}$ crosses zero at ~0.2 wt%, defining an acoustic damping threshold that marks the crossover from temperature-controlled viscoelastic losses to structure-controlled acoustic attenuation, similar to behavior observed in poly(propylene glycol) systems [48, 49].

Finally, local heating softens the matrix more strongly at low Si loading, reducing acoustic velocities relative to global heating. Increasing Si content homogenizes T and elastic response, and at 2 wt% $\Delta_{B\ Shift}$ approaches zero, identifying an elastic homogenization threshold where the nanocomposite responds as a mechanically uniform medium at GHz frequencies.

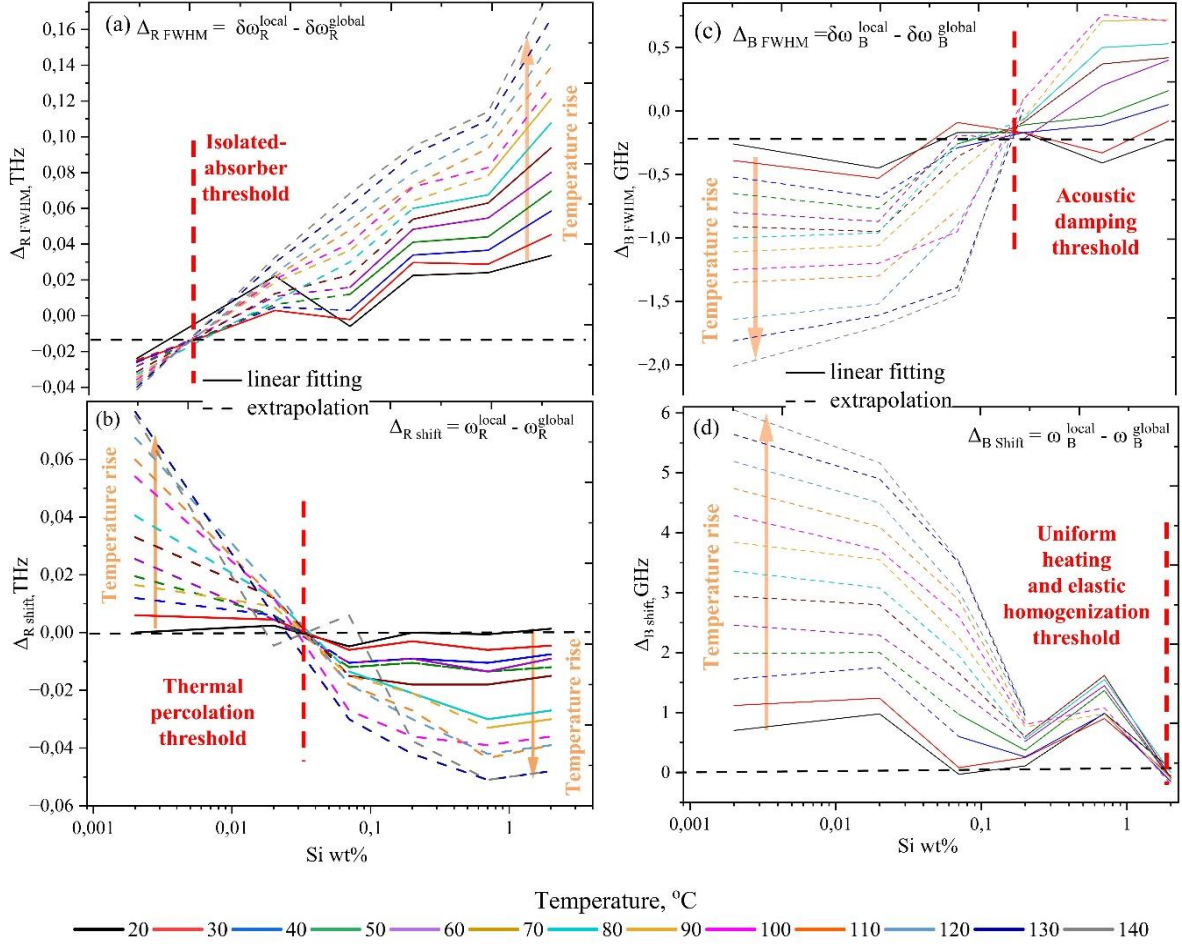

FIG.6. (a, b) The difference in the Raman Si-Si peak position ($\omega_R$) and linewidth ($\delta\omega_R$) between local and global heating methods at different concentrations and heating temperatures, (c, d) The difference in the Brillouin frequency shift ($\omega_B$) and linewidth ($\delta\omega_B$) between local and global heating methods at different concentrations and heating temperatures. Solid lines represent the linear fitting of measured results, while dashed lines show extrapolation of linear fitted data. See Supplementary materials for details of extrapolation.

Figure 7(a) shows Brillouin intensity increasing with T for all concentrations, followed by a peak and sharp collapse. This anomaly's position and sharpness depend on Si concentration. The initial rise reflects enhanced thermal phonon population and elasto-optic coupling from matrix softening. The maximum corresponds to optimal scattering where viscoelastic relaxation meets the GHz Brillouin period ($\omega\tau^B \approx 1$)[28, 50]. The subsequent collapse indicates phonon overdamping from viscous losses. Higher Si loading delays and smoothens this by modifying relaxation. Figure 7(b) shows a monotonic Brillouin frequency decrease with T, with acoustic softening in the highlighted region [51]. Higher Si content yields higher frequencies, indicating a greater effective longitudinal modulus. Figure 7(c) shows the linewidth peaking near ~180 °C, corresponding to maximal GHz attenuation from α-relaxation. Increased Si loading reduces and broadens this peak. Figure 7(d) reveals a DSC glass transition at ~60–70°C, with a smaller heat-capacity step at higher loadings indicating restricted mobility [52]. A degradation transition appears at ~360°C.

The results separate thermodynamic and high-frequency mechanical transitions. The DSC glass transition (~60–70°C) relates to segmental motion [52], while BLS anomalies near ~180°C mark a viscoelastic crossover ($\omega \tau^B \approx 1$) with maximal damping. Si NPs enhance stiffness, suppress damping, and broaden this crossover via interfacial confinement. Degradation at ~360°C is slightly delayed by Si NPs.

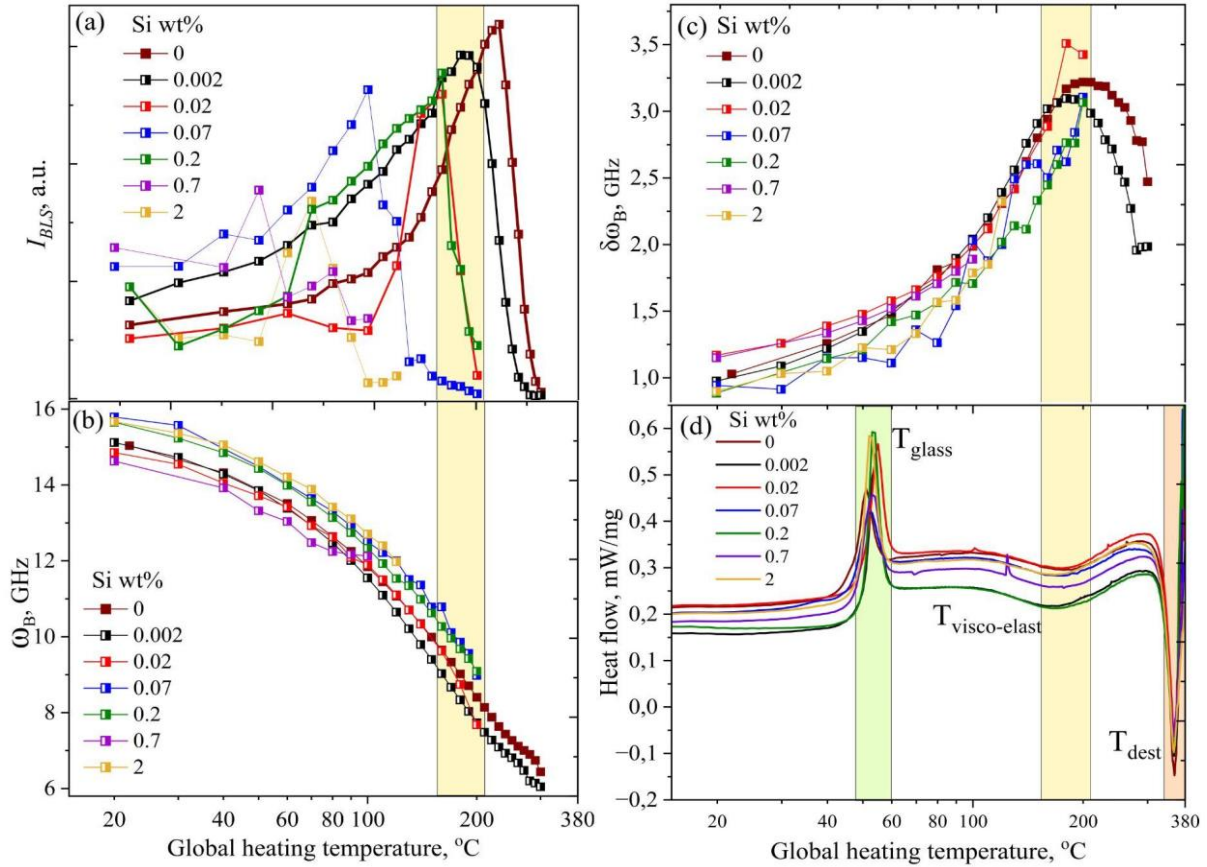

FIG. 7. Measured Brillouin peak intensities (a), shifts (b) and linewidths (c), as well as DSC measured heat flow rate per unit mass versus *global* heating temperatures at various concentrations.

In summary, combined Raman and Brillouin micro-spectroscopy reveals a decoupling between local and global thermo-phononic responses in nano-Si/epoxy composites. Raman-derived isolated absorber (~ 0.003 wt%) and thermal percolation (~ 0.07 wt%) thresholds are controlled by optical absorption and heat diffusion. Brillouin-derived acoustic damping (~ 0.2 wt%) and uniform heating (~2 wt%) thresholds are governed by elastic homogenization and viscoelastic relaxation. Increasing T preserves this threshold sequence, defining a progression from isolated nanoscale hot-spots to a uniform composite.

Raman spectroscopy shows optical phonon softening and broadening under local heating at low Si loadings, consistent with isolated nanoheaters. Overlapping thermal fields at higher content mark thermal percolation. Brillouin scattering reveals elastic softening and a linewidth maximum near ~200 °C, corresponding to a GHz α-relaxation crossover. Higher Si loadings shift acoustic attenuation toward interfacial scattering and polymer confinement, signaling elastic homogenization.

Compared to SiC NWs, Si NP fillers demonstrate better heat conductive properties of resulting epoxy-based nanocomposites at 2 wt% filler loading. However, despite bulk silicon's high conductivity, heat transport in nano-Si/epoxy composites remains matrix-dominated and interface-limited, while phonon lifetimes are modified by NPs. Si NPs act as interface-engineered phonon-damping and viscoelastic-modifying agents, reshaping epoxy's frequency-dependent response. This decouples the glass transition from the GHz mechanical crossover, stabilizing coherent acoustic phonons into the rubbery regime.

These findings establish dual spectroscopy as a powerful framework for probing nanoscale hot-spot formation and viscoelastic dissipation, with practical relevance to vibration damping, thermal shock resistance, and self-diagnostic materials.

**Supplementary material**

See the **Supplementary Material** for detailed information about the assessment of light and heat penetration depth values, and data interpolation process.

**Acknowledgements**

This work is supported by 111024CRP2003 and 20122022CRP1608 grants via Collaborative Research Program (CRP) and 20122022FD4130 grant via Faculty Development Competitive Research Grants Program (FDCRGP) of Nazarbayev University and by AP19679332 grant from Kazakhstan Ministry of Science and Higher Education.

**Conflict of interest**

The authors declare no conflict of interest.

**Data availability**

The raw data required to reproduce these findings are available from the corresponding author, upon email request. The processed data required to reproduce these findings are also available upon request.

Supplementary Materials for

# Probing optical and acoustic phonons in heated nano-Si/epoxy composites

Bayan Kurbanova[1,*], Vladimir Bessonov[1], Ivan Lysenko[2], Gauhar Mussabek[3], Ali Belarouci[4], Vladimir Lysenko[5], Yanwei Wang[6,7] and Zhandos Utegulov[1]

[1]*Department of Physics, School of Sciences and Humanities, Nazarbayev University, Astana 010000, Kazakhstan*
[2]*Institute of High Technologies, Taras Shevchenko National University of Kyiv, 64/13 Volodymyrska St., Kyiv 01601, Ukraine*
[3]*Faculty of Physics and Technology, Al-Farabi Kazakh National University, Almaty 050040, Kazakhstan*
[4]*Lyon Institute of Nanotechnology, UMR 5270, INSA de Lyon, 69100 Villeurbanne, France*
[5]*Light Matter Institute, UMR-5306, Claude Bernard University of Lyon/CNRS, Université de Lyon, 69622 Villeurbanne Cedex, France*
[6]*Department of Chemical and Materials Engineering, School of Engineering and Digital Sciences, Nazarbayev University, Astana 010000, Kazakhstan*
[7]*Center for Energy and Advanced Materials Science, National Laboratory Astana, Astana 010000, Kazakhstan*

*Corresponding author*: bayan.kurbanova@nu.edu.kz

Figures S1 and S2 illustrate spectra behaviours for *local* and *global* heating regimes at various Si wt%, respectively. The Raman optical phonon frequency decreases and Raman linewidth increases with the temperature rise for all Si loadings. At a given temperature, samples subjected to *local* laser heating exhibit a stronger red shift and peak broadening compared to *global* heating, particularly at higher NP concentrations. The Brillouin frequency decreases and Brillouin linewidth increases with temperature, corresponding to thermal softening of elastic moduli and enhanced acoustic phonon attenuation. Under *global* heating (FIG.S2) the maximum measured temperature at which Raman signals were still detectable was 140 ºC. However, the equivalent *local* laser-heating temperature was not attainable due to low optical absorption in samples with low Si loading and due to burning in samples with high Si loading due to excessive optical absorption. Therefore, in order to predict missing local temperatures and enable their comparison with global temperatures, a linear extrapolation was employed to extrapolate (by dashed lines) beyond the fitted (experimental results are linearly fitted) accessible data (displayed by connected solid lines), as shown in FIG. S1.

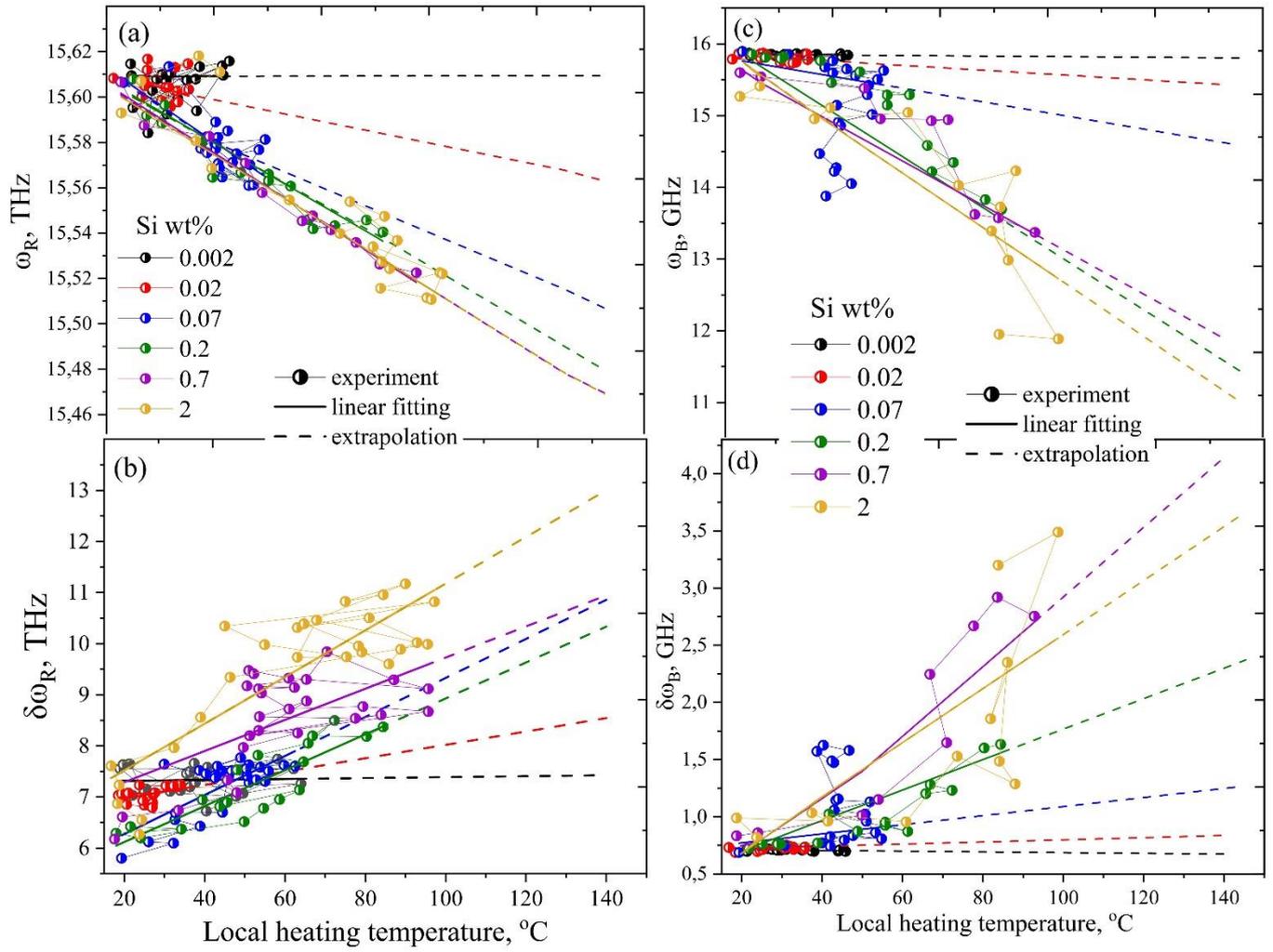

FIG. S1. Measured shifts and linewidths of Raman (a ,b) and Brillouin peaks (c, d) for *local* heating temperatures for various Si NP concentrations. Points represent measured data, solid lines - interpolated lines and dashed lines - extrapolated lines in the range of temperatures not accessible spectroscopically

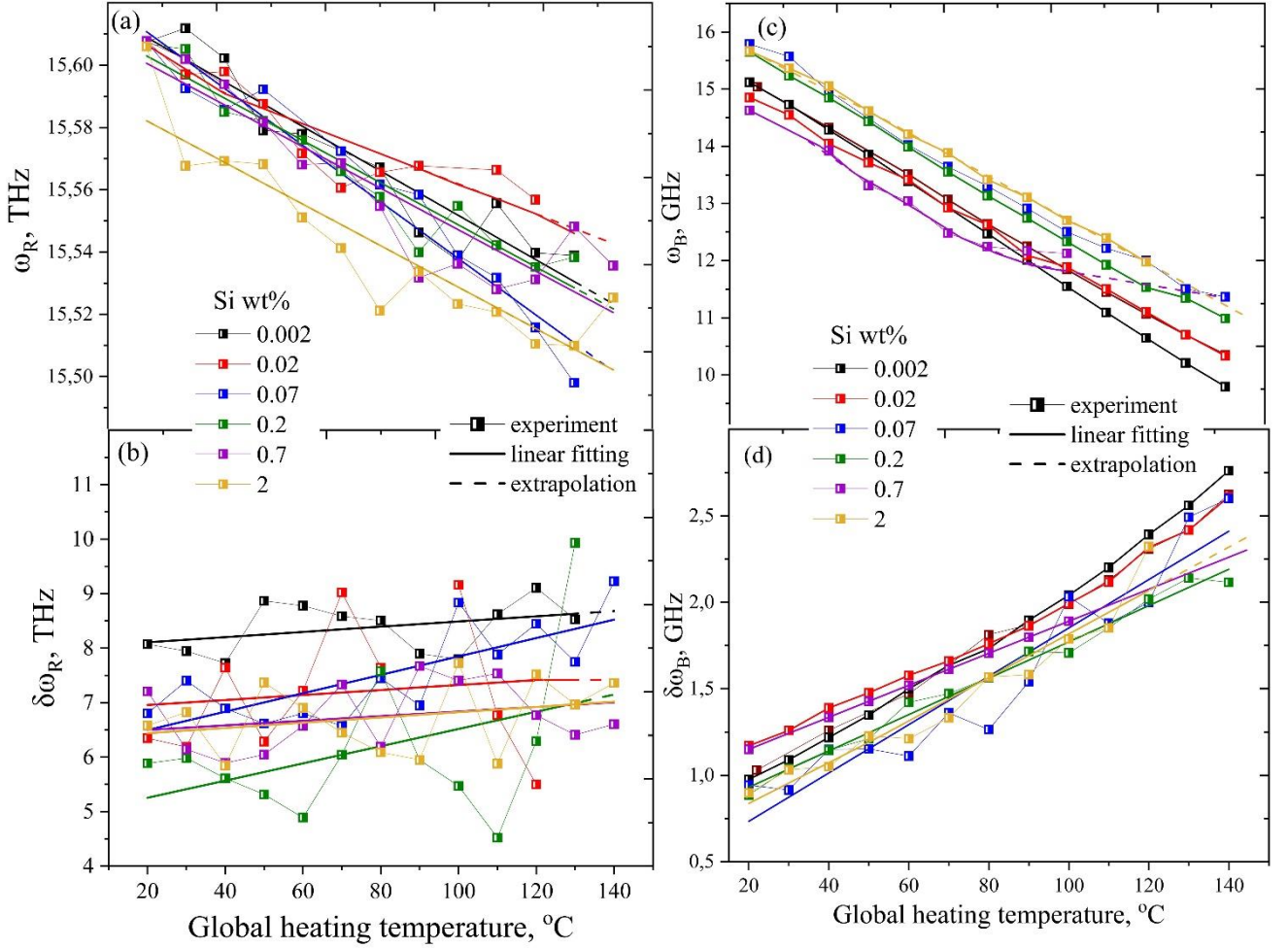

FIG. S2. Measured shifts and linewidths of Raman (a ,b) and Brillouin peaks (c, d) vs *global* heating temperatures for various Si NP concentrations. Points represent measured data, solid lines - interpolated lines and dashed lines - extrapolated lines in the range of temperatures not accessible spectroscopically

The following formula was used to calculate the light penetration depth into the composite material [SM1]:

$$l_{penetration} = \frac{\lambda_0}{4\pi n_2} \quad (S1)$$

where $n_2$ is the imaginary part of the effective refractive index of the composite material, $\lambda_0$ is the wavelength of light in a vacuum. $n_2$ is related to the effective permittivity by the equation [1]:

$$n_2 = \sqrt{\frac{Abs(\epsilon_{eff}) - Re(\epsilon_{eff})}{2}} \quad (S2)$$

The effective permittivity was calculated using Maxwell's effective medium theory, since the filling factor ($f_{Si}$) in our case does not exceed 2%, allowing us to limit the analysis to a linear approximation:

$$\epsilon_{eff} = \epsilon_m + 3 f_{Si} \epsilon_m \frac{\epsilon_{Si} - \epsilon_m}{\epsilon_{Si} + 2\epsilon_m} \tag{S3}$$

where the coefficients "m" and "Si" denote the epoxy matrix and silicon, respectively. The filling factor $f_{Si}$ of Si NPs is directly determined by their concentration $c_{Si}$ according to the formula:

$$f_{Si} = \frac{c_{Si}}{\rho_{Si}} \tag{S4}$$

where $\rho_{Si}$ is the density of Si.

The initial values of density and specific heat capacity of the composite material are calculated using the following formulas:

$$\rho = f_{Si}\rho_{Si} + (1 - f_{Si})\rho_m$$
$$C = (f_{Si}\rho_{Si}C_{Si} + (1 - f_{Si})\rho_m C_m)/\rho \tag{S5}$$

The total absorbed light power density is calculated using the formula:

$$P_{total} = \rho_{density} \frac{\pi d^2}{4} 10^{-3} \text{ W} \tag{S6}$$

Table S1 summarizes the calculated laser light (532 nm) penetration depth and thermal conductivity values for np-Si/epoxy composites with varying Si NP concentrations measured by opto-thermal Raman thermometry. The light penetration depth depends inversely with NP loading: at high Si content (2 wt%), the penetration depth is very small (0.17 mm), indicating strong optical absorption and highly localized energy deposition near the surface. As the Si concentration decreases, the penetration depth increases rapidly, reaching 0.5 mm at 0.7 wt%, 1.7 mm at 0.2 wt% and 5.1 mm at 0.07 wt%. For samples having low Si loading (0.02 and 0.002 wt%), due to very weak optical absorption the optical penetration depth increases dramatically to 17 mm and 168 mm, respectively. For these two samples having nearly bulk-transparent behavior it was not possible to determine thermal conductivity values using FEM.

The steady-state Fourier heat conduction equation is solved numerically using the laser light penetration depth values to determine thermal conductivity values. The resulting photo-induced spatial temperature distribution within the bulk and and on the surface of the composites is calculated in correlation with the dependence of heating temperature versus absorbed light power density measured by Raman thermometry (FIG. 3a), using the thermal conductivity of the composite materials as the sole fitting parameters. The error values in the thermal conductivity were determined from variations in the slope of the temperature versus absorbed light power density curves.

The figure S3 shows simulated cross-sectional temperature distributions (x–z plane) under laser heating, illustrating how the thermal profile evolves with increasing NP concentration from panels (a) to (d). Overall, the rise of Si NP loading causes higher peak temperatures and increasingly localized heating near the surface, accompanied by a pronounced reduction in thermal penetration depth. At low concentrations, heat spreads more deeply into the bulk, whereas at high concentrations, enhanced optical absorption produces strong, surface-confined photothermal hot spots while the bulk remains near ambient temperature.

Table S1. Laser light (532nm) penetration depth and thermal conductivity values for different concentrations

| Si wt% | 2 | 0.7 | 0.2 | 0.07 | 0.02 | 0.002 |
|---|---|---|---|---|---|---|
| Light penetration depth, mm | 0.17 | 0.5 | 1.7 | 5.1 | 17 | 168 |
| Thermal conductivity, W/mK | 0.46±0.07 | 0.21±0.04 | 0.12±0.01 | 0.09±0.03 | Not available due to low absorption | Not available due to low absorption |

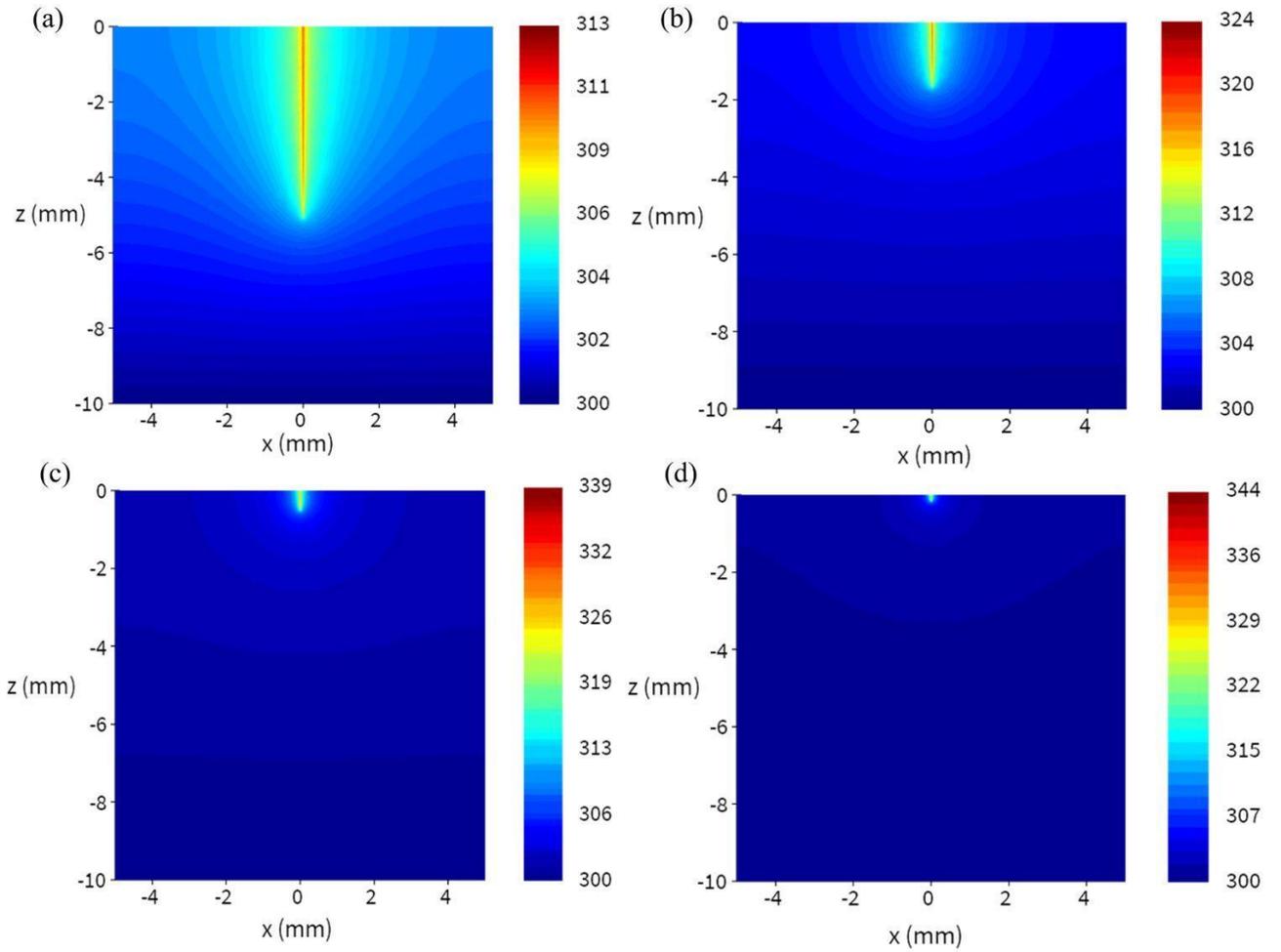

FIG. S3. Cross-sectional laser-induced local temperature distributions in the x–z plane at various penetration depths under laser heating for various NP loadings and corresponding heat penetration depths: (a) 0.07 Si wt%, $z_{therm}$ = 5.5 mm , (b) 0.2 Si wt%, $z_{therm}$ = 2 mm, (c) 0.7 Si wt%, $z_{therm}$ = 0.5 mm, (d) 2 Si wt%, $z_{therm}$ = 0.2 mm.

References
[SM1] E. Hecht, *Optics* (Addison-Wesley, 2002)